\documentclass[preprint,review,1p,times,12pt]{elsarticle}

 \usepackage{graphicx}

\usepackage{amssymb}
\usepackage{amsmath,bm}

\journal{Comptes Rendus - Physique}

\begin{document}

\begin{frontmatter}



\title{Mechanisms and origin of multiferroicity}


\author{Paolo Barone}
\ead{paolo.barone@spin.cnr.it}

\author{Silvia Picozzi}
\ead{silvia.picozzi@spin.cnr.it}

\address{Consiglio Nazionale delle Ricerche (CNR-SPIN), 67100 L'Aquila, IT}

\begin{abstract}
Motivated by the potential applications of their intrinsic cross-coupling properties, the interest in multiferroic materials has constantly increased recently, leading to significant experimental and theoretical advancements. From the theoretical point of view, recent progresses have allowed to identify different mechanisms responsible for the appearence of ferroelectric polarization coexisting with -- and coupled to --  magnetic properties. This chapter aims at reviewing the fundamental mechanisms devised so far, mainly in transition-metal oxides, which lie at the origin of multiferroicity.
\end{abstract}


\end{frontmatter}



\section{Introduction}

Magnetic and ferroelectric materials, characterized by a spontaneous symmetry breaking that causes the appearance of a switchable
long-range magnetic or dipolar ordering below a critical temperature, are ubiquitous in modern science and present-day
technology for their diverse applications, ranging from data storage to magnetic and ferroelectric
random-access memories\cite{ziese_mag, magnetism_handbook, scott_fe, rabe_topics}.
In recent years, the quest for magnetoelectric multifunctional
integration within a single material has motivated a renewed interest in the class of so-called
multiferroic materials, displaying the simultaneous presence of two or more spontaneous ferroic
phases\cite{schmid_fe}. The intrinsic combination of magnetism and ferroelectricity in this class of compounds
calls for novel device paradigms exploiting their cross-coupled effect, e.g.
allowing to control magnetization (polarization) by an external electric (magnetic) field. The
intense research activity on this class of materials, both from the experimental and theoretical
sides, is testified by the number of excellent review papers on the topic, opening promising
prospects for applications also beyond the field of magnetic and ferroelectric materials, e.g. for
information and energy harvesting technologies\cite{SpaldinFiebig_2005, cheong_rev, ramesh_rev, Khomskii, wang_rev, tokura_rev, pyatakov_rev}.

From the theoretical point of view, the intrinsically coupled and multifunctional nature of
multiferroics poses numerous fundamental challenges. In fact, if on the one hand the theoretical
understanding of magnetic insulators is rather well established, on the other hand a rigorous
microscopic theory for ferroelectrics has been formulated only in the last few decades.
The microscopic origin of magnetism is basically the same in all magnetic insulators, being related to the presence of localized
electrons, mostly in the partially filled $d$ or $f$ shells of
transition-metal or rare-earth ions, which have a corresponding
localized magnetic moment. Exchange interactions between the localized moments, usually accounted for in general Heisenberg-like spin models,
lead to magnetic order\cite{stohr}. For ferroelectrics, the situation is quite different, and one major difficulty comes from the fact that 
electric dipoles in extended-state systems are well defined only in the very special case of a fully ionic material,
while in general the electric polarization is a global property of the matter that cannot be
decomposed in localized contributions. A rigorous definition of ferroelectric bulk polarization has been provided
only in the early 90s via a quantum-mechanical approach based on Berry phases\cite{Vanderbilt1, Vanderbilt2, Resta}, which may explain the difficulties encountered in identifying possible microscopic mechanisms responsible for ferroelectricity. In fact, the microscopic origin of ferroelectricity is not unique, and the coexistence and
interplay between structural distorsions and electronic degrees of freedom (spin, charge, and
orbital) has been proven to play a central role in the diverse mechanisms devised so far. In this
respect, first-principles calculations based on the density-functional theory (DFT) have played
an important role in the description, understanding as well as prediction of magnetic, ferroelectric and magnetoelectric properties
of mutiferroics, due to its ability to describe the many active degrees of freedom within a
comparable level of accuracy\cite{PicozziEderer_2009, alesilvia_rev, stroppa_picozzi_2010} (see also the chapter of Ghosez and collaborators).

When discussing the origin and microscopic mechanisms for multiferroicity, therefore, the main problem lies in the origin of ferroelectricity, its coexistence with and its coupling to the magnetic ordering. Generally speaking, one can identify two kind of mechanisms
leading to the appearence of ferroelectric polarization. The first is
substantially ionic/displacive, in the sense that it involves lattice distortions
of ions carrying different charges, where the structural deformation lies at the basis of the
space-inversion symmetry breaking. The second kind of mechanisms, on the other hand, involves primarily
electronic degrees of freedom, which are responsible for the space-inversion
symmetry breaking of the electronic ground state even in centrosymmetric
structures. Such a classification is to many extents an approximation,
since the two mechanisms are tightly intertwined and a signature of symmetry-lowering
in both ionic and electronic sectors is usually found in the very same
material. A somewhat related, but possibly less arbitrary, distinction is that between proper and improper ferroelectricity:
in improper ferroelectrics, the spontaneous polarization can be considered as a by-product of another structural (or electronic) primary phase transition, as opposed to proper ferroelectrics, where the symmetry lowering can be ascribed primarily to polar distortions. In the following, we will review the microscopic mechanisms for multiferroicity that have been proposed and identified so far, paying special attention to the understanding of the origin of ferroelectricity and its coexistence/coupling with magnetic ordering.

\section{Microscopic mechanisms for proper multiferroics}

\subsection{Conventional ferroelectric perovskites: revisiting the exclusion rule}
Most of the long-known (conventional) ferroelectrics are perovskite
transition-metal oxides (BaTiO$_3$, PbTiO$_3$, KNbO$_3$,  Pb(ZrTi)O$_3$) in which the polar distortion
mainly involves an off-centering of the perovskite B-site transition-metal cation showing an
empty $d$ shell, which is prone to establish some degree of covalency with surrounding oxygens\cite{Matthias, Lines77}.
On the basis of a
lattice shell-model, Cochran\cite{cochran_advph_1960, cochran_advph_1961} showed that covalent interactions tend
to be short range, while the ionic electrostatic interactions are long (in fact infinite) range.
The competition between short-range forces, which tend to
favour high-symmetry phases, and long-range Coulomb interactions, which
destabilize the centrosymmetric structure, is then influenced by the strong
covalency that softens the cation-O repulsion, leading to an offcentering of
the transition-metal ion towards one (or three) oxygen(s), at the expense of
weakening the bonds with other oxygen ions.
In a simplified local approach, the microscopic origin of the cation offcentering can be deduced by estimating the energy associated to the deformation of the covalent bonds in a ligand-cation-ligand geometry as 
\begin{eqnarray}
\delta E\simeq -\frac{t_{pd}^2}{\Delta}(1+gu)^2-\frac{t_{pd}^2}{\Delta}(1-gu)^2+2\frac{t_{pd}^2}{\Delta}=-2\frac{t_{pd}^2}{\Delta}(gu)^2,
\end{eqnarray}
where $t_{pd}$ is the hopping matrix element describing hybridization interactions between the cation and the O-$p$ states, $\Delta$ is the charge-transfer gap, $u$ describes the distortion and the changes in hybridization are accounted for as $t_{pd}(u)\simeq t_{pd}(1\pm gu)$ in the linear approximation\cite{khomskii_jmmm_2006}. If the elastic energy cost of the lattice distortion, $\sim K u^2/2$, is smaller than the quadratic covalency energy gain, then the local distortion is energetically favourable, whereas the different charges associated to different ions may explain qualitatively the appearence of local electric dipoles\footnote{Due to the covalent character of the bonds, however, a quantitative estimate of bulk ${\bm P}$ is only accessible via the Berry-phase formulation in the framework of the modern theory of polarization; in this context, each ion carries dynamical Born effective charges (BECs), instead of nominal static ones, and the strong covalent effect is usually reflected in anomalously large BECs.}. 

A more refined (but still substantially local) approach is based on the vibronic coupling between ground and excited electronic states, whose expression can be deduced by general symmetry considerations, that is known to trigger (pseudo) Jahn-Teller distortions\cite{bersuker_pl_1966,bersuker_book}. In this approach, the adiabatic potential energy surface near the high-symmetry configuration contains a vibronic contribution $K=K_0+K_\nu$. $K_0$ is a positive contribution that coherently includes the local odd displacements of all the atoms, which is a long-range
(whole crystal) feature, whereas the matrix elements of $K_\nu$, which contains negligible intercell interaction terms and is found to be always negative, are nonzero when the overlap
between the wave functions of the ground state of
atoms of one sublattice (oxygens) with the excited state of
the atoms of the other sublattice (transition-metal cations)
increases due to the nuclear displacements $u$, thus enhancing
the covalency. In this framework, the microscopic mechanism  leading to ferroelectric transition in a crystal is directly related to {\it both its atomic and electronic structures}, where the origin of polarization appears to be substantially local but depends strongly on long-range interactions.

Besides providing an example of the intrinsically coupled nature of ferroelectricity in crystal systems, the case of conventional perovksites also allows to introduce one of the major theoretical challenge that has been faced in the field of magnetoelectric multiferroics, i.e. the understanding of which conditions must be met in order to combine magnetism and ferroelectricity in a single homogeneous material. In fact, for a long time the two phenomena have been tought to be mutually exclusive and chemically incompatible in perovskite oxides, due to the requirement of having empty (partially filled) $d$ states in order to have ferroelectricity (magnetism)\cite{hill_jpcb_2000,filippetti_prb_2002}. However, a careful analysis based on the pseudo Jahn-Teller effect has shown that such exclusion rule is less strict than expected even in the class of proper ferroelectric
oxides, where some specific $d^n$ and spin configurations can allow, on the basis of symmetry-allowed (lattice-electron) vibronic couplings, for the coexistence of magnetism and ferroelectricity\cite{bersuker_prl_2012,bersuker_chemrev_2013,bersuker_conf_2013}.
A particularly interesting case
is that of high-spin $d^3$ configuration, a situation realized in alkaline-earth manganites such as CaMnO$_3$ or SrMnO$_3$;
these systems were shown to possess a weak ferroelectric instability mediated by a covalency-driven mechanism, which at ambient conditions is hindered by other energetically favourable distorsions, namely
nonpolar rotation/tiltings of BO$_6$ octahedra \cite{ghosez_prl,rondinelli_prb, barone_prb}. Applying strain or chemical pressure could in principle tune the balance between different structural instabilities, allowing for
ferroelectricity to emerge, as indeed recently shown for strained CaMnO$_3$ films\cite{cmo_film} and
bulk Sr$_{1-x}$Ba$_x$MnO$_3$\cite{sakai}. Since both magnetic and ferroelectric instabilities are related to the
same manganese B cations, the magnetoelectric coupling is expected to be strong in these
systems\cite{lee_rabe, giovannetti_SBMO_prl}, even though antiferromagnetic phases are likely to appear; on the
other hand, critical temperatures may be far larger than in frustrated magnets (e.g. a $T_c\sim 185 ~K$ has been reported for Sr$_{0.5}$Ba$_{0.5}$MnO$_3$\cite{sakai}).

\subsection{Ferroelectricity due to lone pairs}
The recent boost in the multiferroic field can be reasonably ascribed also to the successful realization of BiFeO$_3$ thin films with enhanced multiferroic properties  by the Ramesh's group\cite{Wang_Science}.
This perovskite oxide, one of the most celebrated multiferroic materials due both to its very large ferroelectric polarization and to its high ferroelectric as well as magnetic critical temperatures, apparently escaped from the aforementioned exclusion rule. However, it turned out that the microscopic origin of ferroelectricity in this system is completely different, not involving polar offcentering of transition-metal B-site cations; instead, the main instability was related to the presence of stereochemically active lone pairs on the A-site Bi ions\cite{neaton, goffinet}. The ferroelectricity arises here from on-site $sp$ rehybridization of the two $6s$ electrons of bismuth that do not participate in chemical
bonds, thus showing a high polarizability whose particular orientation may create local dipoles and trigger the onset of the observed very large ferroelectric polarization ($P\sim 100~\mu C/cm^2$\cite{expBFO}). It is interesting to note that the presence of nonpolar tilting/rotation of the BO$_6$ octahedra, that ultimately seems to prevent the appearence of ferroelectricity in orthorhombic alkali-metal manganites, is not competing with the ferroelectric instability in BiFeO$_3$, where polar distortions involve primarly A-ions; on the contrary, a primary role of the BO$_6$ rotational patterns has been recently suggested in the ferroelectric phase of this compound\cite{dieguez}. 
On the other hand, magnetism is guaranteed by the B-site magnetic Fe$^{3+}$ ($d^5$) ions. Due to the independent origin of ferroelectricity and magnetism in this
type of multiferroics, one could have expected the magnetoelectric coupling to be rather small; however, a number of recent observations indicate that the magnetoelectric coupling in BiFeO$_3$ can
be significant and lead to unusual and very interesting effects, suggesting also an important role played by the large ferroelectric polarization
in determining a long-period modulation of the magnetic ordering\cite{kadomtseva, park, johnson, tokunaga}.

Beside BiFeO$_3$, the lone-pair mechanism could be realized in other related systems. Indeed, it was first theoretically proposed for BiMnO$_3$\cite{seshadri}, even though later computational and experimental structural optimizations revealed that the ferroelectric phase is not stable and suggested that BiMnO$_3$ belongs
to the centrosymmetric $C2/c$ structure\cite{baettig, Belik07}. The ferroelectric (or antiferroelectric) nature of this material is still debated, as several details -- such as oxygen stochiometry -- seem to play a crucial role in determining both the structure and the properties of BiMnO$_3$\cite{sundaresan}. Recently, it has been also proposed that BiMnO$_3$ could behave as an improper ferroelectric, where ferroelectricity could develop from a specific antiferromagnetic ordering which would break the inversion symmetry even in a centrosymmetric crystal structure\cite{solovyev1,solovyev2}. On the other hand, a similar mechanism has been suggested to be at play in the relatively new multiferroic PbNiO$_3$\cite{Inaguma, Inaguma-2, Hao, Hao2}, despite the fact that the nominal valence of Pb is $4+$, thus no lone pairs should be active on A-ions; DFT-based results have however clarified that the Pb formal valence is perturbed by Pb $6s$ - O $2p$ hybridization which results in enhanced ferroelectric instability and predicted large polarization ($P\sim 100 \mu C/cm^2$).

\section{Microscopic mechanisms for improper multiferroics}

\subsection{Hybrid improper ferroelectricity}
In the quest for general mechanisms which may give rise to both ferroelectric and magnetic orderings induced by the same lattice instability, a significant theoretical advance came from a careful reconsideration of rotational distortions in transition-metal ABO$_3$ perovskites. Octahedron rotations are in fact ubiquitous in perovskites and related materials, and are known to strongly couple to magnetic properties, because they buckle the inter-octahedral B-O-B bond angles which mediate the interplay of electronic (spin and orbital) degrees of freedom\cite{ImadaRMP}. Although they are usually found to compete with ferroelectric instabilities in conventional ferroelectric oxides, it has been recently shown that some combination of such rotations can induce local polar displacements of A-site cations through effective force fields exerted by O ions, even though the distortional patterns that are most frequently found in simple perovskites result in cooperative antipolar (antiferrodistortive) distortions\cite{jorge_bellaiche}. However, this observation suggests the exploration of related systems where the effect could be exploited to engineer an improper ferroelectric state arising from suitable combinations of proper rotational distortions (which are individually nonpolar). The key ingredient is a trilinear coupling between a polar mode $P$ (that individually would be energetically unfavourable) with two unstable nonpolar lattice modes $R_1, R_2$ with different symmetries, whose combined symmetry properties $R_1\oplus R_2$ are the same 
of the polar one. The mechanism, that was originally proposed to explain ferroelectricity in a Aurivillius layered compound\cite{perez-mato_hybrid}, has been called ``hybrid improper ferroelectricity'', highlighting the improper origin of ferroelectric polarization and the role of coupled nonpolar distortions\cite{benedek_fennie}. The fact that the polarization is proportional to the product of two distortions implies that it
will flip its direction if either $R_1$ or $R_2$, though not both, is reversed. Interestingly, the trilinear coupling may also serve as the interlink between magnetic and ferroelectric properties, i.e. provide a mechanism for linear magnetoelectric effects. This has been thoroughly discussed in Ref. \cite{benedek_fennie}, where a layered perovskite Ca$_3$Mn$_2$O$_7$ served as a model system to illustrate the coupling between ferroelectricity. The theoretical magnetic ground state displays a canted antiferromagnetic configuration with a weak ferromagnetic moment, arising from a relativistic Dzyaloshinskii-Moriya interaction that is strongly dependent on the buckling of Mn-O-Mn angles. Interestingly, a linear magnetoelectric effect was predicted to appear, mediated by the rotational modes which are responsible at the same time for the ferroelectric polarization and the (weak) ferromagnetic moment. 
Since these nonpolar distortions usually manifest near or above room temperature, there appears to be no fundamental limitation on the temperature range over which this mechanism could give rise to multiferroic order, thus motivating an intense research activity aiming at identifying suitable design rules to single out promising candidates, as discussed in the chapter by Ghosez et al.

\begin{figure}
\includegraphics[width=0.95\textwidth]{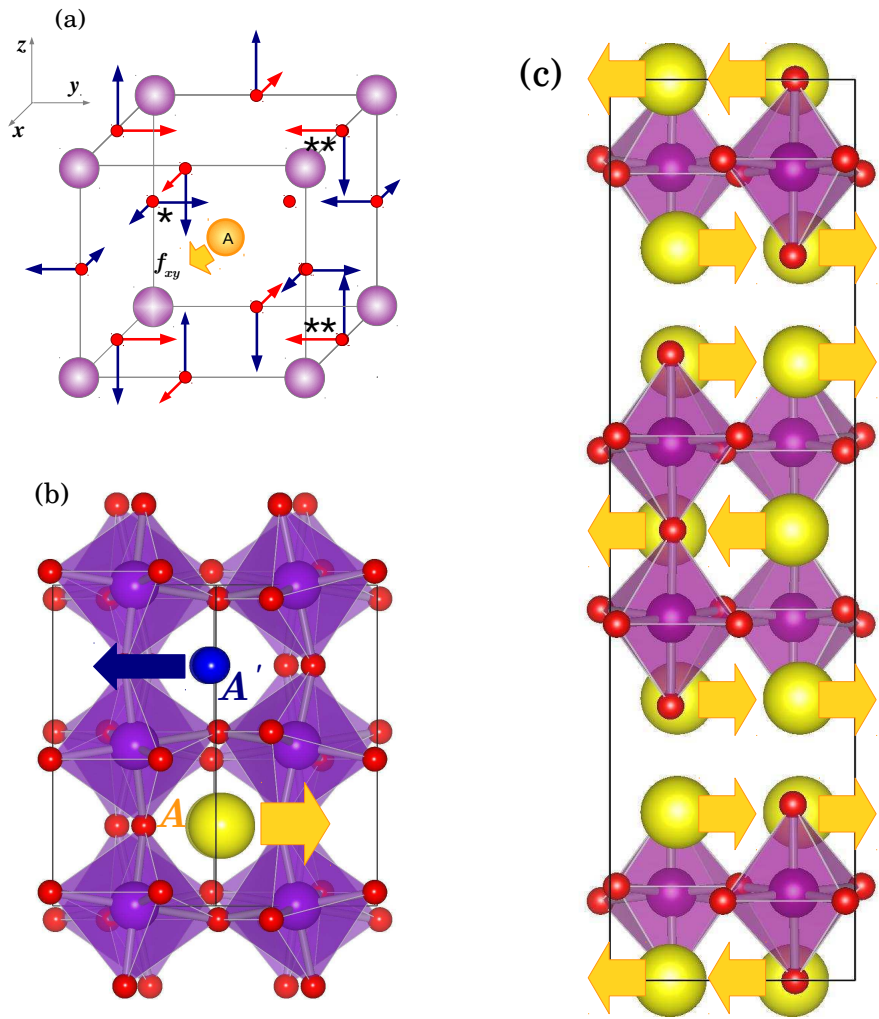}
\caption{Hybrid improper ferroelectricity arising from a cooperative interplay of rotation and tilting distortions as usually found in the $Pnma$ orthorhombic structure. (a) The distorted oxygen positions are responsible for a net force field acting on $A$-site ions (asterisks mark the oxygens that get closest to the central A-cation). In the $Pnma$ structure, characterized by an $a^-a^-c^+$ pattern in Glazer's notation, the $A$ ions undergo antipolar displacements, which usually do not give rise to a net electric polarization. When such antipolar displacements are not compensated, as in the case of ordered layers of $A$ and $A'$ ions (b) or in the layered structure found in the Ruddlesden-Popper phases (c), a bulk polarization can develop from the interplay of the nonpolar rotational distortions.}
\label{fig:hyb}
\end{figure}

The mechanism has been identified in several materials that are potential improper ferroelectric and multiferroic, including perovskite artificial superlattices\cite{bousquet_nat,jorge_superlattices}, layered (ABO$_3$)$_2$(AO)\cite{benedek_fennie} or double AA'BB'O$_6$ perovskites with specific A/A' cation orderings\cite{fuku_stroppa, rondinelli_advmat}. Interestingly, it has been recently shown how a very similar mechanism can be active at twin walls naturally occurring in many ABO$_3$ centrosymmetric perovksites (such as CaTiO$_3$ or PbZrO$_3$)\cite{paolo_preprint}, where polar displacements have been predicted\cite{salje_prl} and then observed to occur at the wall\cite{cto_TEM,pzo_TEM}. Due to the structural origin of such effect, a wall polarization is expected to develop also at twin walls of magnetic nonpolar oxides, such as CaMnO$_3$ or LaFeO$_3$, that would be possibly coupled to the magnetic ordering. Eventually, other nonpolar lattice modes could be included beside rotations and tilting; as a matter of fact, hybrid improper ferroelectricity and magnetoelectric effects - in principle allowing for an electrical control of magnetization -  have been predicted in related metal-organic frameworks with perovskite structure, where the nonpolar mode mediating both effects is a Jahn-Teller distortion of oxygen octahedra coupled with a tilting mode\cite{stroppa_angchem, stroppa_advmat}. As opposed
to their inorganic counterparts, metal-organic frameworks offer the possibility of varying their
additional degrees of freedom, arising from organic/inorganic duality, and are currently the object of an increasing interest as potential novel multiferroic materials\cite{viart_mof}.

\subsection{Magnetically induced ferroelectricity}
Among the class of improper multiferroics, materials displaying magnetically-induced ferroelectric polarization have been object to an intense research activity in the last decade, due to their large intrinsic magnetoelectric coupling\cite{cheong_rev,wang_rev,tokura_rev,pyatakov_rev,kimura_rev}. Generally speaking, the ferroelectric polarization in these materials arises as an improper (or secondary) transition induced by a specific magnetic ordering, which is itself responsible for the symmetry lowering and loss of the space inversion even on top of a centrosymmetric structure. The way by which the magnetic ordering breaks the inversion symmerty is two-fold, since the symmetry breaking may occur in the magnetic electronic ground state (even in centrosymmetric crystal lattices) or may be caused by structural distortions induced by specific configurations of the magnetic moments (a mechanism generally called exchange striction). 

At the basis of the first kind of mechanisms lies the intrinsic charge anisotropy of individual $d$ orbital states, that is ultimately responsible for both covalent effects and exchange interactions in transition-metal oxides, whereas the coupling between spin and orbital degrees of freedom can be mediated by the atomic spin-orbit coupling or by correlation effects such as the Hund coupling.
In this respect, model approaches have been devised to identify possible mechanisms by which a local electric dipole may develop from electronic spin degrees of freedom, typically in a cluster model taking into account the transition-metal orbital states and their hybridization with surrounding ligand ions, mediating the effective $d$-$d$ hopping processes and crystal-field splittings\cite{knb, jia_prb_2006, jia_prb_2007, hu_prb_2008, whangbo_prl_2011, whangbo_prb_2013,yamauchi_prb_2011,khomskii_prb_2008}.  
Under general assumptions (substantially related to the time-reversal invariance), it is possible to show that the electric dipole of a spin dimer with inversion symmetry can be always written in a cluster expansion approach as:
\begin{eqnarray}\label{eq:general_spin}
\bm P =\bm P_l (\bm S_l)+\bm P_r (\bm S_r)+ \bm p_{lr}\bm S_l\times\bm S_r,
\end{eqnarray}
where the local contributions must be a quadratic form of the corresponding local spin, $\bm P_a=\sum_{\alpha,\beta} \bm \pi_a^{\alpha\beta}S_{a\alpha}S_{a\beta}$, $a=l,r$ and $\alpha,\beta$ labeling the cartesian components $x,y,z$\cite{whangbo_prl_2011}, while $\bm p_{lr}$ is a 3 x 3 matrix. Assuming some microscopic model for the cluster, the explicit dependence of the coefficients $\bm \pi_a$ and $\bm p_{lr}$ on microscopic quantities can be derived. The easiest way is to consider two transition-metal ions interacting via some ligand, described by a model Hamiltonian $H=H_L+ H_{t} + H_{SOC} + H_{TM} + H_U$, where $H_L$ ($H_{TM}$) describes the local $p$ ($d$) states on the ligand (transition-metal) sites, $H_t$ accounts for the electron transfer through the ligands, which is described by an hybridization matrix depending on the $p,d$ orbitals involved and on their relative position, while $H_{SOC}$ describes the spin-orbit coupling on the transition-metal ions. The last term $H_U=-U\sum_{a=l,r}\hat{\bf m}_a\cdot{\bf S}_a$ describes a local Zeeman-like term which allows to tune the spin-dimer configuration. In a three-ion linear cluster with the ligand occupying the middle of the bond, two contributions to polarization have been predicted\cite{knb, jia_prb_2006, jia_prb_2007,hu_prb_2008}, arising from the mixing of $d$ orbitals induced by the spin-orbit coupling and hybridization effects mediated by the ligand oxygen. This mixing triggers the onset of a local dipole in the cluster, intimately connected to the shapes of involved $d$ and $p$ orbitals. 
From the nonzero overlap integral $\langle d_{a,yz}\vert {\bf y} \vert p_z \rangle$, a transverse component
of the local electric dipole is found, 
\begin{eqnarray}\label{eq:knb}
{\bf P} \propto {\bf e} \times {\bf S}_l\times{\bf S}_r,
\end{eqnarray}
where $\bf e$ is the unit vector parallel to the bond direction\cite{knb,jia_prb_2007}. Following Katsura, Nagaosa and Balatsky, the electric dipole is said to be induced by a spin-current mechanism, since the
vector product ${\bf S}_l\times{\bf S}_r$ is proportional to the spin current ${\bf j}_s$, where the
Dzyaloshinskii-Moriya vector $\bf d$ acts as its vector potential\cite{knb}. Within the same cluster geometry, a purely electronic longitudinal component of the electric dipole also arises from the nonzero overlap integrals $\langle d_{a,xy}\vert {\bf x}
\vert p_y \rangle$ and $\langle d_{a,zx}\vert {\bf x} \vert p_z \rangle$. This contribution to $\bf P$ results from the
variation of the $d$-$p$ hybridization mediated by the spin-orbit coupling, with a predicted functional form in the linear three-ion cluster\cite{jia_prb_2006,jia_prb_2007}:
\begin{eqnarray}\label{eq:bondp}
{\bf P} \propto ({\bf S}_l\cdot{\bf e})\,{\bf S}_l-({\bf S}_r\cdot{\bf e})\,{\bf S}_r
\end{eqnarray}
showing its local character (involving only single-spin terms) as opposed to the previously discussed mechanisms for inter-site spin-dependent hybridization effect. From general symmetry argument, it can be shown that the only non-zero coefficients entering  in Eq. (\ref{eq:general_spin})
are $\pi^{xx}_l=(C_1, 0, 0)$, $\bm \pi^{xy}_l=\pi^{yx}_l=(0,C_1/2,0)$, $\pi^{xz}_l=\pi^{zx}_l=(0,0,C_1/2)$ and $\bm p_{lr}(2,3)=-\bm p_{lr}(3,2)=C_2$ in the linear three-ion model, and Eq. \ref{eq:general_spin} reduces to the former equations  (\ref{eq:knb}), (\ref{eq:bondp})\cite{whangbo_prl_2011}.
However, the explicit expressions of the coefficients $\bm p_{lr}$ and $\bm \pi_a$ depend on the specific geometry of the cluster and/or the material under considerations, affecting local electronic structure, crystal-field splittings and hopping interactions. For instance, according to the conventional spin-current mechanism of Eq. (\ref{eq:knb}), the helical spin-spiral configuration in triangular-lattice antiferromagnetic CuFeO$_2$\cite{kimura_prb_2006,seki_prb_2007} or MnI$_2$\cite{kurumaji_prl_2011} should not support any ferroelectric polarization. On the other hand, a combined DFT and cluster model approach has shown that indeed a generalized spin-current mechanism can be active in the specific triangular-lattice geometry. In fact, the presence of two ligands connecting the magnetic ions is responsible for the appearence of other nonzero terms in the $\bm p_{lr}$ matrix, allowing for the appearence of a polarization even in the helical spin configuration\cite{whangbo_prl_2011}. On the other hand, the local contribution to $P$ arising from the spin-dependent hybridization mechanism in tetrahedral crystal fields, as found in melilite crystals, has been shown to behave as $\bm P=\sum_a (\bm S\cdot\bm e_a)^2\,\bm e_a$, where $a=1,..,4$ labels four oxygens surrounding the central magnetic ion and $\bm e_a$ are unit vectors parallel to the metal-oxygen bonds\cite{murakawa_prl_2010,yamauchi_prb_2011}, whereas a more complex dependence have been found in Cu$_2$OSeO$_3$, displaying CuO$_5$ trigonal bipyramids and square pyramids\cite{whangbo_prl_2012}. 

\begin{figure}
\includegraphics[width=0.98\textwidth]{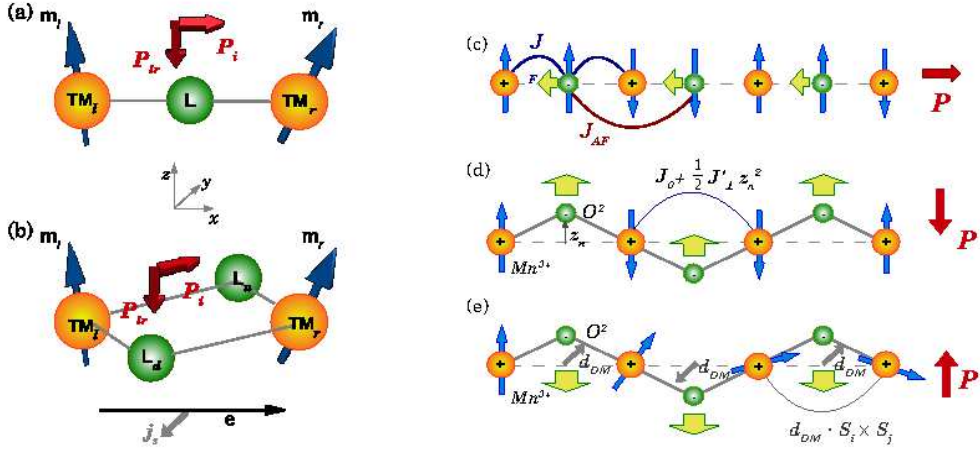}
\caption{(a) and (b) Two possible choices of cluster models describing a spin dimer with one or two bridging ligand ions. If inversion symmetry is preserved, two contributions to local electric dipole are expected, $\bm P_i$ depending only on the local spin configuration on each transition-metal site and $\bm P_{lr}$ generally being proportional to $\bm S_l\times \bm S_r$. In a linear three-ion model (a), the intra-site contribution gives rise to a purely longitudinal contribution to polarization, whereas the spin current induces a transverse local dipole. In the four-site cluster (b), the longitudinal $\bm P_i$ develops in the plane of the cluster, its direction depending on the spin-dimer configuration; on the other hand, the spin-current polarization is not always perpendicular to the spin current, having different in-plane and out-of-plane components. (c) Magnetostrictive effect in the  $\uparrow\,\uparrow\,\downarrow\,\downarrow$ Ising model. (d), (e) Magnetostrictive effects in Mn-O chains representative of $R~$MnO$_3$, with strong oxygen buckling distortions. The spins on each Mn ion experience a superexchange coupling $J\simeq J_0+\frac{1}{2}J~'_\perp z_n^2 >0$, which is invariant under inversion symmetry and depends only on even powers of $z_n$; assuming $z_n=(-1)^n z_0+\delta z_n$,
small O displacements are energetically favourable if $\delta z_n=(-1)^n (J~'_\perp z_0/\kappa) \cos\theta_n$, where
$\theta_n$ is the
angle between neighbouring spins and $\frac{\kappa}{2}\sum_n \delta z_n^2$ is the elastic energy associated to the distortion.
When considering the antisymmetric Dzyaloshinskii-Moriya interaction, the staggering of the Dzyaloshinskii vector along the chain implies that the corresponding magnetic energy is minimized by $\delta z_n\propto(\lambda/\kappa)\sin\theta_n$; in this case, a collective O shift is allowed only by a spiral
spin configuration, since the vector product $\bm S_n\times\bm S_{n+1}$ has the same sign for all
pairs of neighboring spins
}\label{fig:spin}
\end{figure}

Even though a unified cluster-based approach that attempts to describe both pure electronic and ion-displacement contributions has been recently proposed\cite{whangbo_prb_2013}, magnetostrictive effects have been usually described in the general context of spin-model Hamiltonians, where the effect of structural distortions on exchange coupling constants has to be suitably included, depending on the material under consideration. A necessary condition for magnetically-induced polar distortions is that the magnetic ordering itself breaks the inversion symmetry, i.e. only inhomogenous or frustrated spin systems can support a polar displacement. Microscopically, lattice distortions in the presence of competing magnetic interactions
can then appear in order to maximize the gain in magnetic energy. A simple example is provided by an Ising-spin chain with competing ferromagnetic nearest-neighbour and antiferromagnetic next-nearest-neighbour interactions, where an inhomogeneous $\uparrow\,\uparrow\,\downarrow\,\downarrow$ spin configuration can be realized. If some energy is gained, e.g., by the
shortening of a ferromagnetic bond, the chain will distort in such a way that bonds between parallel spins shorten
and those between antiparallel ones stretch, causing polar displacements that would result in a net bulk polarization if different charges are carried by each site along the chain. The simplest realization of this magnetostrictive mechanism can be found in Ca$_3$CoMnO$_6$, where Co$^{2+}$ and Mn$^{4+}$ magnetic cations alternate along one-dimensional chains\cite{choi_prl_2008,wu_prl_2009}. 
On the other hand, the (super)exchange processes in transition-metal oxides
are mediated by the ligand oxygens bridging the metal cations, and they generally depend on ionic relative positions.
Since oxygens carry nominal negative charge (as opposed to the positive one carried by magnetic ions) their collective shift can
induce a finite polarization. Different magnetic configurations can be responsible for different polar displacements through the symmetric or antisymmetric exchange interaction, as discussed heuristically in Ref.\cite{sergienko_prb_2006} for rare-earth manganites $R$MnO$_3$. In order to maximize the magnetic energy gain, the Mn-O-Mn bond angle mediating the exchange interactions may be increased or decreased, causing a polar displacements of oxygens.  A possible spin ordering leading to polar displacement
is the $\uparrow\,\uparrow\,\downarrow\,\downarrow$ configuration, where the Mn-O-Mn angle increases (decreases) at ferro- or antiferromagnetic alternating bonds, leading to a collective displacement of ligand O driven by the symmetric exchange interaction. On the other hand, the antisymmetric  (or Dzyaloshinskii-Moriya) exchange interaction is responsible for a polar magnetostriction if a cycloidal spin configuration sets in in the orthorhombic structure of rare-earth manganites, as shown in Fig. \ref{fig:spin}. The polarization driven by such inverse Dzyaloshinskii-Moriya mechanism appears to have the same dependence on spin configuration as the spin-current polarization derived before; however, the latter has been derived in a frozen ionic configuration and describes the purely elecronic contribution to the ferroelectric polarization, whereas the former describes a magnetostrictive mechanism where ionic displacements are induced mainly by the magnetic energy optimization. In principle, both an electronic and an ionic mechanism can contribute to the total polarization, even though 
it is not easy to predict which contribution is larger.

\subsection{Ferroelectricity due to charge/orbital ordering}
Improper ferroelectricity is also in principle allowed in materials displaying specific charge or orbital orderings which break the inversion symmetry, individually or in combination with a magnetic ordering. Charge ordering is a rather ubiquitous effect in transition-metal
oxides, being often observed in materials with ions having formally a mixed valence like Fe or Ni. It is also likely to appear
in low-dimensional charge-transfer organic salts. Generally speaking, different forms of charge ordering may appear.
Typically in materials with rather localized electrons, the electronic charge disproportionation takes place mainly on top
of ions, leading to a so-called site-centered charge ordering. On the other hand, the Peierls transition often observed, e.g.,
in quasi-one-dimensional organic salts is characterized by a bond-centered charge-density
wave, implying that charge disproportionation takes place roughly on alternating strong
and weak bonds between ions. Each time a given charge-ordered configuration breaks the space-inversion
symmetry  --- for instance, in the presence of a mixed site-centered and bond-centered charge-density wave, as proposed in \cite{efremov} --- a macroscopic polarization is likely to appear\cite{khomskii_brink.rev}. There are, however, no general rules for the realization
of either one situation or the other (or a combination of the two), and charge ordering effects have to be considered separately
case by case. The simplest situation is realized when charge-ordered ions occupy sites belonging to a structure that lacks inversion symmetry, as in the case of some of the proposed low-temperature structures of magnetite Fe$_3$O$_4$\cite{kuni_magnetite} or in the double-layer structure of LuFe$_2$O$_4$, displaying Fe$^{2+}$-rich and Fe$^{3+}$-rich layers\cite{nat_lufeo,whangbo_lufeo}. When taking into account the covalency, such ionic picture can be modified and a purely electronic contribution to polarization can appear; this effect can be understood in the framework of the Berry-phase theory of polarization by looking at the Wannier-function centers, describing the center of charge of the continuous electronic charge density, which in the presence of covalent effects may move from centrosymmetric positions (site-centered or bond-centered), thus leading to the appearence of local dipoles\cite{kuni_paolo.rev}.

\begin{figure}
\includegraphics[width=0.98\textwidth]{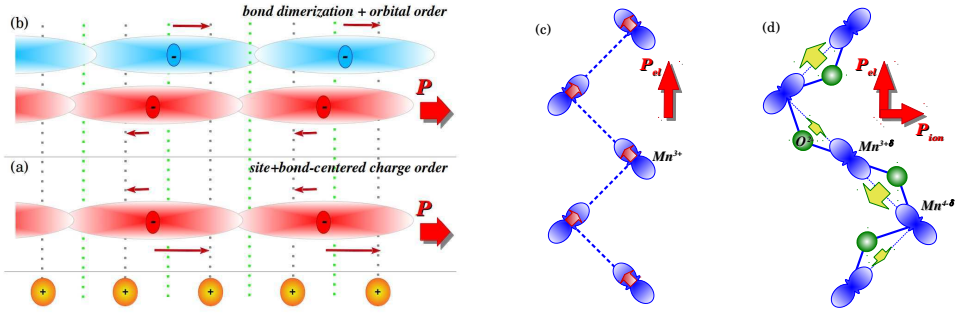}
\caption{(a) and (b) Sketch of a neutral one-dimensional chain displaying a mixed site-bond charge-ordering wave (a) or an orbital ordering in the presence of bond-dimerization (b). In both cases the Wannier-function centers are displaced with respect to inversion-symmetric positions, highlighted by vertical dotted lines, thus giving rise to uncompensated local electric dipoles and, hence, to a macroscopic polarization. (c) Schematic representation of electronic contribution to polarization expected in the E-type antiferromagnetic configuration found in some undoped manganites, such as HoMnO$_3$; electronic hopping is constrained by strong Hund's rule only along one-dimensional zig-zag chains in MnO$_2$ layers, and is further asymmetrized by the onset of orbital ordering triggered by Jahn-Teller and correlation effects. The $e_g$-like Wannier-function centers are then coherently displaced from the Mn sites, giving rise to a Berry-phase polarization parallel to the chain direction\cite{barone_etype}. (d) Schematic diagram of the electronic polarization and buckling-induced mechanism predicted in the CE-type antiferromagnetic configuration observed in half-doped manganites, such as Pr$_{0.5}$Ca$_{0.5}$MnO$_3$; the interplay of correlation effects and buckling distortion is responsible for orbital ordering and a bond-dimerization along each ferromagnetic chain, where the Mn-O-Mn angle between Mn ions with parallel occupied orbitals is increased more than the angle between Mn ions with perpendicular orbitals. Polar displacements of bridging oxygen cause the appearance of an ionic $P$ perpendicular to the chain, while orbital ordering leads to an electronic Berry-phase $P$ parallel to the chain direction\cite{barone_ce}.
}\label{fig:co_oo}
\end{figure}

Similarly, orbital ordering may also cause the appearence of electronic contribution to polarization when the electronic charge density acquires a noncentrosymmetric distribution that can be also traced by the positions of the corresponding Wannier-function centers\cite{kuni_paolo.rev}. In its simplest realization, orbital-induced polarization can appear when an orbital ordering develops in a system with bond dimerization and inequivalent hopping integrals; the orbital-ordered state can then be viewed as a superposition of two inequivalent charge-localization phenomena in different
orbital sectors, as shown in Fig. \ref{fig:co_oo}b), leading to uncompensated local electronic dipoles which sum up to a bulk polarization\cite{barone_oo}. This mechanism could be realized, e.g., in vanadate spinels such as CdV$_2$O$_4$, showing both Peierls-like bond dimerizations and an ordering of $d_{yz}, d_{zx}$ orbitals. A different scenario has been proposed for undoped and half-doped manganites\cite{barone_etype, soluyanov_oo, barone_ce}, whose electronic properties can be properly described
in the limit of infinite Hund coupling, which induces an infinite intra-atomic splitting between minority- and majority-spin
states. The relevant low-energy model is a degenerate double-exchange Hamiltonian describing $e_g$ spinless electrons
whose motion is determined by the underlying magnetic configuration of the almost localized $t_{2g}$ spins. The combination of E-type antiferromagnetic ordering and Jahn-Teller lattice distortions, as found in undoped manganites, remove the orbital degeneracy leading to an ordered state with alternating $d_{3x^2-r^2}/d_{3y^2-r^2}$ along zig-zag chains of parallel spins in each MnO$_2$ layer. In turn, this leads to a change of the Berry phase associated to 
the alternating right-handed and left-handed motion of the $e_g$ electrons around each site displaying the orbital ordering, which is ultimately related to the appearence of an electronic polarization\cite{barone_etype, soluyanov_oo}. As for the case of spin-induced electronic polarization, the mechanism relies on the coupling between spin and orbital degrees of freedom, where the magnetic configuration determines the way in which $d$ states hybridize and electrons hop in the lattice and a local dipole develops depending on the shape of individual orbitals and their
hybridization through oxygen's $p$ orbitals; in this case, however, the strength of such coupling is determined by Hund's interaction, rather than the weaker spin-orbit coupling, with a consequent larger value of the expected contribution to polarization (as indeed predicted by DFT calculations\cite{picozzi_etype, colizzi_ce}).

\section{Conclusions}
In this chapter, we have presented a tutorial introduction to the microscopic mechanisms of multiferroicity that have been unveiled so far. The recent theoretical advances in the field seem to suggest that the multiferroic phenomenology is much more common than what originally believed. The reason probably lies in the variety of mechanisms leading to ferroelectricity and to its coexistence with other kind of orderings, involving also electronic degrees of freedom. Even though several theoretical approaches have been adopted to identify the origin of multiferroicity in its diverse forms, the strength of the microscopic mechanisms and their efficiency in leading to large polarization can be most likely assessed by resorting to first-principles techniques, as thoroughly discussed in the chapter by Ghosez and collaborators. Nonetheless, the intuitive understanding hopefully provided in this chapter might provide the basic knowledge required for the search and optimization of multiferroic properties in novel materials beyond the class of perovskite oxides, including even more complex oxides, such as materials with spinel or melilite structure, or organic-inorganic hybrid systems as the recently proposed metal-organic frameworks.





\bibliographystyle{model1a-num-names}
\bibliography{barone_picozzi_biblio.bib}







\end{document}